\documentclass[twoside]{article}
\usepackage{fleqn,espcrc2}
\usepackage{graphicx}

\usepackage{graphicx}

\newcommand{\AmS}{{\protect\the\textfont2
  A\kern-.1667em\lower.5ex\hbox{M}\kern-.125emS}}

\title{Chemical potential response of meson masses at finite temperature}

\author{
{\sl QCD-TARO Collaboration:}
 Ph.~de~Forcrand\address{Inst. f\"ur Theoretische Physik,
             ETH-H\"onggerberg, CH-8093 Z\"urich, Switzerland}
             \address{CERN, Theory Division, CH-1211 Geneva 23, Switzerland},
 T.~Hashimoto\address{Department of Applied Physics, Faculty of Engineering,
             Fukui University, Fukui 910-8507, Japan},
 S.~Hioki\address{Department of Physics, Tezukayama University,
                Nara 631-8501, Japan},
 Y.~Liu\address[DP]{Department of Physics, Hiroshima University,
                  Higashi-Hiroshima 739-8526, Japan},
 H.~Matsufuru\address{Research Center for Nuclear Physics, Osaka University,
            Ibaraki 567-0047, Japan},
 O.~Miyamura\addressmark[DP],
 A.~Nakamura\address{Research Institute for Information Science and Education,
             Hiroshima University, Higashi-Hiroshima 739-8521, Japan},
 T. Takaishi\address{Hiroshima University of Economics, Hiroshima 731-01, Japan}
 and T.~Umeda\addressmark[DP]
}
       
\begin{document}

\begin{abstract}

We study the response of meson masses to the chemical
potential ($\partial{m}/\partial{\mu}$) at high temperature and at zero
chemical potential on $N_f=2$ lattice with staggered fermions.
Preliminary results for the meson composed of different quarks
show that 
$\partial{m}/\partial{\mu}|_{\mu=0}$ is negative in the confinement phase and 
 positive in the deconfinement phase.

\vspace{1pc}
\end{abstract}

\maketitle

\section{INTRODUCTION}

The temperature and density effects to the properties of hadrons are
interesting and important for the early universe 
and for high energy physics.
So far the temperature effect has been extensively studied \cite{LAT}.
But there are not so many results concerning the density effect.
As is well known, there are difficulties
with the simulation of a finite density system by the lattice QCD approach.
The reason is that the fermionic determinant becomes complex and
cannot be considered as a part of the probability.  
If we factor out the phase of the determinant and put it 
into observables,
it leads to oscillating contributions in quantum averages.  
This makes it hard to obtain reliable results.
It has been also known that the naive quenched approximation leads 
to an essentially different world \cite{Stephanov}.

In spite of this difficult situation, density effects to hadrons
such as a mass-shift are very interesting
and important subjects both theoretically and
experimentally \cite{Hatsuda-Lee,Agaki}.
There are several approaches to circumvent
this difficulty and they seem successful to a limited extent
\cite{Glasgow}.  
Another way is to study two-color QCD 
\cite{Nakamura84,Lombardo99a,Lombardo99,Morrison,Montvay,Sinclair}.
Also, a non-trivial quenched approximation can be
defined \cite{Cano} by taking the simultaneous limit where 
the quark mass and the 
logarithm of the chemical potential both become infinite \cite{ben}.  
As for the response to the chemical potential, only the baryon number susceptibility
at zero baryon density
has been studied, and an abrupt jump at the transition point has been
reported \cite{Gott}.

In this paper, we examine the response to the chemical potential
of the mass of hadrons at finite temperature and at
zero baryon density in full SU(3) QCD simulations\cite{TARO2}.

\section{CHEMICAL POTENTIAL RESPONSE OF HADRONIC MASSES 
AT HIGH TEMPERATURE}

As stated in the introduction, the lattice study at finite density is still
difficult.  In this situation, an interesting possibility
 is to examine the
response of physical quantities to the chemical potential
at zero chemical potential.  

Here we consider a hadronic correlator $G(x)$,

\begin{equation}
G(x)=\sum_{y,z,t}<H(x,y,z,t)H(0,0,0,0)^{\dagger}>,
\end{equation}
where $H(x,y,z,t)$ denotes the hadron operator.

Suppose that this hadronic correlator
is dominated by a single pole, then
\begin{equation}
G(x) \approx A e^{-mx}.
\end{equation}
We take derivative with respect to chemical potential $\mu$,
\begin{eqnarray}
B(x) &\equiv& G(x)^{-1}{\partial G(x) \over \partial \mu}  \label{deri-1a} \\
  &=& A^{-1}{\partial A \over \partial \mu}
- {\partial m \over \partial \mu} x \ \ .
\label{deri-1}
\end{eqnarray}
In eq.(\ref{deri-1}), if the left hand side is measured as a function of $x$,
the linear term gives the chemical potential response of the
hadron mass, while the constant term gives the response of the coupling.

The next problem is  to get the derivative of the correlator. For this
purpose, we go back to the definition of the hadronic correlator.  

\begin{eqnarray}
& &<H(n)H(0)^{\dagger}>= \nonumber \\
& & Z^{-1} \int [dU] \mbox{Tr}\,(M(n;0)\Gamma M(0;n)\Gamma^{\dagger})\nonumber \\
& & \mbox{det}(D)\exp(-S_G)
\label{deri0}
\end{eqnarray}
where
$Z=\int [dU] \mbox{det}(D)\exp(-S_G)$ 
 and  $M = D^{-1}$.

Then, using the formulae
\begin{eqnarray}
{\partial \mbox{det}(D) \over \partial \mu} =
\mbox{Tr}\,(M {\partial D \over \partial \mu}) \mbox{det}(D)
\end{eqnarray}
and
\begin{eqnarray}
{\partial M \over \partial \mu}
= - M{\partial D \over \partial \mu}M ,
\end{eqnarray}
we get
\begin{eqnarray}
& & {\partial<H(n)H(0)^{\dagger}> \over \partial \mu} = \nonumber \\
 \ \ \ \ &-& <\mbox{Tr}\,(M\dot{D}M\Gamma M\Gamma^{\dagger})> \nonumber \\
 \ \ \ \ &-& <\mbox{Tr}\,(M\Gamma M\dot{D}M \Gamma^{\dagger})>  \nonumber \\
 \ \ \ \ &+& <\mbox{Tr}\,(M\Gamma M\Gamma^{\dagger}) 
              \mbox{Tr}\,(\dot{D}M)>\nonumber \\
 \ \ \ \ &-& <\mbox{Tr}\,(M\Gamma M\Gamma^{\dagger})> <\mbox{Tr}\,(\dot{D}M)>
\label{deri2}
\end{eqnarray}
where the short hand notations 
\begin{eqnarray}
\dot{X} = {\partial X \over \partial \mu}
\end{eqnarray}
and
\begin{eqnarray}
< Y > = Z^{-1} \int [dU] \mbox{det}(D)\exp(-S_G) Y
\end{eqnarray}
are used. Here we calculate the first and second terms of 
the right hand side of eq.(\ref{deri2}) and 
ignore the other terms.
$\mbox{Tr}\,(\dot{D}M)$ is pure imaginary and its average value,
i.e. $ <\mbox{Tr}\,(\dot{D}M)>$ 
will be zero\cite{Gott}.
If $\mbox{Tr}\,(M\Gamma M\Gamma^{\dagger})$ is real, the third term 
does not contribute as real and we may neglect it.  
For degenerate quarks,  $\mbox{Tr}\,(M\Gamma M\Gamma^{\dagger})$ is real.
Although we consider here non-degenerate systems
for which  $\mbox{Tr}\,(M\Gamma M\Gamma^{\dagger})$ is not real,
we assume that the contribution from the third term is small and 
drop it in this study.  

We introduce two independent chemical potentials, $\mu_q$ and $\mu_Q$
for two different quarks ( $m_q \leq m_Q$ ).
We introduce  convenient definitions of the chemical potential,
$\mu_s \equiv (\mu_q + \mu_Q )/2$  and $\mu_v \equiv (\mu_q - \mu_Q )/2$ 
and define the following derivatives with respect to $\mu_s$ and $\mu_v$.
\begin{equation}
\frac{\partial}{\partial \mu_s}=\frac{\partial}{\partial \mu_q} 
+\frac{\partial}{\partial \mu_Q}
\label{deris}
\end{equation}

\begin{equation}
\frac{\partial}{\partial \mu_v}=\frac{\partial}{\partial \mu_q} 
-\frac{\partial}{\partial \mu_Q}
\end{equation}
$\mu_s$ is usual chemical potential corresponding to baryon number.
In this study we consider the response with respect to $\mu_s$.

In the case of the staggered fermion formalism, the derivative of the fermion
operator is
\begin{eqnarray}
& &{\partial D_{KS} \over \partial \mu} = \nonumber \\
& &{a \over 2}[ U_t(n)e^{a\mu}\delta_{n+t,m}\nonumber \\
&+& U_t^{\dagger}(n-t)e^{-a\mu}\delta_{n-t,m}] .
\label{deri3}
\end{eqnarray}
At $\mu=0$, we can evaluate eqs. (\ref{deri0}) and (\ref{deri2}) by usual
techniques of lattice QCD simulation.

\section{PRELIMINARY RESULTS}

Here we present our preliminary result for $\partial{m}/\partial{\mu}|_{\mu=0}$.

\subsection{Lattice parameters}
Lattice parameters are as follows:
\begin{itemize}
\item {\bf Lattice Size}:\\
 $16 \times 8 \times 8 \times 4$, 
with 16 for $x$ direction, 4 for the temperature direction.
\item {\bf Quarks}:\\
 $m_Q$=0.25 and $m_q$=0.025 for $\beta$=5.26,
and 
 $m_Q$=0.25 and $m_q$=0.0125 for $\beta$=5.33.
Dynamical K-S fermions with $N_f=2$ are the light quarks ($m_q$).
The meson operator, $H$, is made of the two different quarks.
\item {\bf Temperature}:\\
 $\beta=5.26$ for $T \approx 0.97T_c$ (confinement), and \\
 $\beta=5.33$ for $T \approx 1.07T_c$ (deconfinement).
\item {\bf Statistics}:\\
 40-90 configurations for each parameter.
\end{itemize}

\subsection{$\partial{m}/\partial{\mu_s}|_{\mu=0}$}

The results for the $B(x)$ in eq.(\ref{deri-1a}) are shown
in Fig.1 and Fig.2.
Fig.1 is for $\beta=5.26$ which corresponds to the confinement phase and
Fig.2 is for $\beta=5.33$ which corresponds to the deconfinement phase.

\begin{figure}[htb]
\includegraphics*[width=20pc]{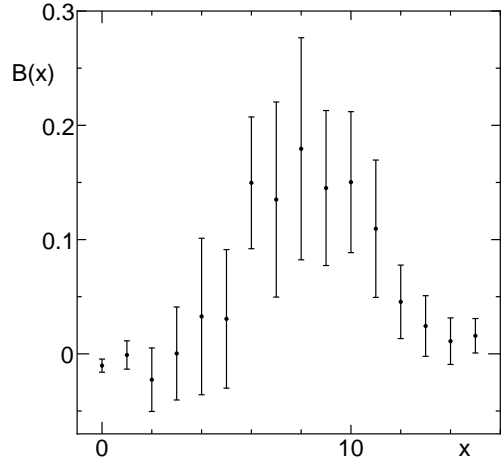}
\caption{ $G(x)^{-1}{\partial G(x) \over \partial \mu_s}_{\mu=0}$
in the confinement phase ($\beta=5.26$, $T \approx 0.97T_c$), 
where $G(x)$ is the hadron propagator
along the $x$ direction.
$m_Q=0.25$ and $m_q=0.025$}
\end{figure}

As is stated above, the coefficient of the linear term in $B(x)$ 
corresponds to $-\partial{m}/\partial{\mu}|_{\mu=0}$.
Although this is preliminary and the statistics are small,
we can see that 
$\partial{m}/\partial{\mu}|_{\mu=0}$ is negative in the confinement phase and
$\partial{m}/\partial{\mu}|_{\mu=0}$ is positive in the deconfinement phase.

\begin{figure}[htb]
\includegraphics*[width=20pc]{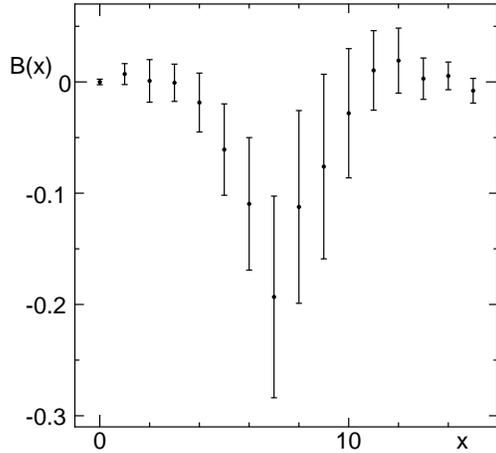}
\caption{ $G(x)^{-1}{\partial G(x) \over \partial \mu_s}_{\mu=0}$
in the deconfinement phase ($\beta=5.33$, $T \approx 1.07T_c$), 
where $G(x)$ is the hadron propagator along the $x$ direction.
$m_Q=0.25$ and $m_q=0.0125$}
\end{figure}

\section{$J/\psi$ SUPPRESSION WITHOUT QGP}

One of our masses, $m_Q$, is in the region of the strange quark.
If the behaviors observed in Figs. 1 and 2 are also
seen in the charm quark mass region, then
the mass of the heavy-light meson, $D$, might
decrease at very small chemical potential just below the
deconfinement temperature.
Then the mass of $D\bar{D}$ system might be lower than
the mass of the $\psi '$ meson,
and a new decay mode
\begin{eqnarray}
\psi ' \rightarrow D + \bar{D}
\end{eqnarray}
will open. 

If this happens in, for example, ultra-relativistic heavy ion collisions,
$\psi '$ mesons produced by this collision will decay as
$\psi ' \rightarrow D + \bar{D}$
and eventually the amount of $J/\psi$ produced through
$ \psi ' \rightarrow \chi \rightarrow J/\psi$ 
channel will decrease.
This means that  $J/\psi$ suppression might occur 
even in the confinement phase as suggested by Hayashigaki\cite{Hayashigaki}.

\section{SUMMARY}

The results of this paper are very preliminary and qualitative
and should be checked in the near future.
However a model calculation based on NJL model\cite{Kuni}
suggests the same tendency.
 
\section*{ACKNOWLEDGMENT}
We thank I. O. Stamatescu for a critical reading of the manuscript.
This work is partially supported by the Grant-in-Aide for
Scientific Research by JSPS and Monbusho, Japan
(No. 10640272, No. 11440080, No. 11694085, No. 11740159 and No. 12554008).

\end{document}